# Locally Nameless Permutation Types*

Edsko de Vries and Vasileios Koutavas

Trinity College Dublin, Ireland

**Abstract.** We define "Locally Nameless Permutation Types", which fuse permutation types as used in Nominal Isabelle with the locally nameless representation. We show that this combination is particularly useful when formalizing programming languages where bound names may become free during execution ("extrusion"), common in process calculi. It inherits the generic definition of permutations and support, and associated lemmas, from the Nominal approach, and the ability to stay close to pencil-and-paper proofs from the locally nameless approach. We explain how to use cofinite quantification in this setting, show why reasoning about renaming is more important here than in languages without extrusion, and provide results about infinite support, necessary when reasoning about countable choice.

## 1 Introduction

If the $\lambda$-calculus is the quintessential functional language, then the $\pi$-calculus is the quintessential concurrent language. Important concepts in the $\pi$-calculus are input and output of channels, denoted $n!c$ and $n?x$, sequential and parallel composition, denoted by $n!c.P$ and $P \mid Q$, and restriction, denoted by $\nu c.P$ (we give a brief summary to the $\pi$-calculus in Section 2).

In a restricted process $\nu c.P$ channel $c$ is *local*, or invisible to the surrounding processes. For this reason, restriction is considered to be a binder, and $\nu c.P$ and $\nu d.({}^{d}/_{c})P$ are alpha-equivalent. However, $P$ can decide to make $c$ visible to a parallel process by outputting it on another channel, as in the following reduction:

$$(\nu c.n!c.P) \mid n?x.Q \rightarrow \nu c.(P \mid ({}^{c}/_{x})Q)$$

We say that $P$ has *extruded* channel $c$.

In order to reason compositionally about $\pi$-calculus processes we can give an operational semantics by means of labelled transition system (LTS) that describes the interaction between a process and its environment. For instance, we can describe the execution of the left process in the example above without reference to its parallel process as

$$\nu c.n!c.P \xrightarrow{(c)n!c} P$$

* This research was supported by SFI project SFI 06 IN.1 1898.

The semantic rule used to derive this transition is

$$\frac{P \xrightarrow{n!c} P' \qquad n \neq c}{\nu c.P \xrightarrow{(c)n!c} P'}\ \textsc{Ext}$$

Notice that channel $c$ is no longer bound in $P'$, because the environment now knows $c$. Rules such as these in which bound names may become free pose difficulties for both the informal and the formal treatment of names.

In informal (pencil-and-paper) proofs the treatment of names is usually based on the Barendregt convention [2, Conv. 2.1.13, p. 26]:

> If $M_1, \ldots, M_n$ occur in a certain mathematical context (e.g. definition, proof), then in these terms all bound variables are chosen to be different from the free variables.

As explained by Bengtson in his encyclopedic thesis [4, Section 3.1.1] this rule is unsound in the presence of a rule such as Ext. For instance, consider the following invalid lemma (read "$c \# P$" as "$c$ not in $P$"):

**Lemma 1 (Wrong).** *If $m \# P$ and $P \xrightarrow{(c)n!c} Q$ then $m \# Q$.*

*Proof.* By induction on the height of $P \xrightarrow{(c)n!c} Q$. In the case for Ext we simply assume that $c \neq m$ by the Barendregt convention.

It is possible to avoid this problem by adding ad-hoc and relation-specific side conditions to the Barendregt convention [18, Conv. 1.4.10, p. 47], but the soundness of such modified rules is difficult to establish. Moreover, the convention must be further modified when additional relations are introduced in which bound names may become free.

Formal approaches to dealing with name binding abound. They differ in how they represent terms, the idiomatic definition of relations (such as the typing relation or transition relation), the amount of infrastructure necessary to deal with names, etc. In an ideal world alpha-equivalent terms are identified so that no reasoning about alpha-equivalence is necessary and the formal proofs follow pencil-and-paper proofs as closely as possible with a minimum of explicit reasoning about names. Two prime candidates for the formal treatment of names at the moment are the nominal approach based on Nominal Logic [16] and used in Nominal Isabelle [19], and the locally nameless representation with cofinite quantification (LNCQ) [1, 5].

Unfortunately, the nominal approach is subject to syntactic restrictions on relations [20] which prohibit rules such as Ext. These syntactic restrictions are necessary in order to maintain soundness, for much the same reasons that the Barendregt convention must be modified. This means that we cannot use the nominal approach to formalize the $\pi$-calculus semantics in the style above. An alternative semantics in terms of so-called *concretions* [14] (sometimes also called *commitments*) is possible but is less common in the process algebra literature, leading to a chasm between pencil-and-paper proofs and formal proofs.

Rules like Ext pose no difficulty for the LNCQ approach. The "cofinite quantification" part of the LNCQ approach refers to a particular style of formalizing relations with the purpose of avoiding the need for renaming as much as possible in proofs. Unfortunately, cofinite quantification turns out to be less effective in the presence of rules such as Ext (Section 6) so that renaming is still frequently required.

Reasoning about renaming is something the nominal approach excels at. In particular, *permutation types* [11] give a generic definition of a type equipped with an operation to apply a (name) permutation and lead to a definition of the *support* of a term (usually, but not always, the set of free names in the term) which is independent of the syntax under consideration. This makes it possible to prove a large set of lemmas which hold across *all* permutation types, reducing the amount of work necessary to formalize new languages.

In this paper we therefore propose to combine the permutation types from Nominal Isabelle with the locally nameless representation. The resulting lightweight formalism makes it possible to formalize the standard $\pi$-calculus semantics with rules such as Ext while staying close to pencil-and-paper proofs, and the use of permutation types streamlines reasoning about renaming. In addition, we make the following contributions:

- We explain where to use cofinite quantification in the formalization of such languages. The surprising result is that there are rules *with* binders that do *not* benefit from cofinite quantification and rules *without* binders that *do* (Section 5).
- In languages without extrusion the use of cofinite quantification means that we need to reason about renaming only in rare cases. We show that this is less true for languages with extrusion (Section 6).
- We extend the theory of permutation types by providing a number of examples of and lemmas about infinite support. Infinite support arises in the study of process calculi when considering infinite traces or terms with countable choice. Moreover, the examples help clarify the difference between support and the set of free names (Section 3).
- We give an axiomatization of the resulting *locally nameless permutation types*, indicate where in the proofs of these axioms we need to reason about finite support, and provide a Coq library containing the axiomatization and many derived lemmas. All the proofs in this paper have been formalized in the accompanying Coq sources[1] (Section 4).

## 2 The $\pi$-calculus

In this section we give a brief description of the $\pi$-calculus and its operational semantics. We do not aim to be complete, only to provide an intuition about the $\pi$-calculus for readers not familiar with it. For a detailed description of the $\pi$-calculus see [14].

[1] see Arxiv ancillary files.

**Syntax**

$$P, Q \quad ::= \quad \texttt{nil} \mid c?z.P \mid c!z.P \mid \nu z.P \mid (P \mid Q) \mid \textstyle\sum_{k \in \mathbb{N}} P_k \mid *P$$

**Labelled Transition System**

$$\frac{}{\langle \Gamma, c\,;\, c!n.P\rangle \xrightarrow{c!n} \langle \Gamma, c, n\,;\, P\rangle}\ \textsc{Out} \qquad \frac{}{\langle \Gamma, c\,;\, c?x.P\rangle \xrightarrow{c?n} \langle \Gamma, c, n\,;\, ({}^{n}/_{x})P\rangle}\ \textsc{Inp}$$

$$\frac{\langle \Gamma\,;\, P\rangle \xrightarrow{\alpha} \langle \Gamma'\,;\, Q\rangle \quad n \# \alpha, \Gamma, \Gamma'}{\langle \Gamma\,;\, \nu n.P\rangle \xrightarrow{\alpha} \langle \Gamma'\,;\, \nu n.Q\rangle}\ \textsc{Res} \qquad \frac{\langle \Gamma\,;\, P\rangle \xrightarrow{c!n} \langle \Gamma, n\,;\, Q\rangle \quad n \# \Gamma}{\langle \Gamma\,;\, \nu n.P\rangle \xrightarrow{(n)c!n} \langle \Gamma, n\,;\, Q\rangle}\ \textsc{Open}$$

$$\frac{\langle \Gamma, c\,;\, P\rangle \xrightarrow{c!n} \langle \Gamma, c, n\,;\, P'\rangle \qquad \langle \Gamma, c\,;\, Q\rangle \xrightarrow{c?n} \langle \Gamma, c, n\,;\, Q'\rangle}{\langle \Gamma\,;\, P \mid Q\rangle \xrightarrow{\tau} \langle \Gamma\,;\, P' \mid Q'\rangle}\ \textsc{Comm-L}$$

$$\frac{\langle \Gamma, c\,;\, P\rangle \xrightarrow{(n)c!n} \langle \Gamma, c, n\,;\, P'\rangle \qquad \langle \Gamma, c\,;\, Q\rangle \xrightarrow{c?n} \langle \Gamma, c, n\,;\, Q'\rangle \quad n \# Q}{\langle \Gamma\,;\, P \mid Q\rangle \xrightarrow{\tau} \langle \Gamma\,;\, \nu n.(P' \mid Q')\rangle}\ \textsc{Close-L}$$

$$\frac{\langle \Gamma\,;\, P_n\rangle \xrightarrow{\alpha} \langle \Gamma'\,;\, Q\rangle}{\langle \Gamma\,;\, \sum_{n \in \mathbb{N}} P_k\rangle \xrightarrow{\alpha} \langle \Gamma'\,;\, Q\rangle}\ \textsc{Sum} \qquad \frac{\langle \Gamma\,;\, P\rangle \xrightarrow{\alpha} \langle \Gamma'\,;\, P'\rangle \quad \mathsf{extr}(\alpha) \# Q}{\langle \Gamma\,;\, P \mid Q\rangle \xrightarrow{\alpha} \langle \Gamma'\,;\, P' \mid Q\rangle}\ \textsc{Par-L}$$

$$\frac{\langle \Gamma\,;\, *P \mid P\rangle \xrightarrow{\alpha} \langle \Gamma'\,;\, *P \mid Q\rangle}{\langle \Gamma\,;\, *P\rangle \xrightarrow{\alpha} \langle \Gamma'\,;\, *P \mid Q\rangle}\ \textsc{Rep}$$

**Fig. 1.** Syntax and Operational Semantics

There are many operational semantics for the the $\pi$-calculus. We use an *environmental* semantics [8] for our study because it gives us a convenient vehicle for explaining our results (in particular in the discussion of cofinite quantification, Section 5). However, our results also apply to other semantics (indeed any other relation) with extrusion-like rules in which bound names become free.

The operational semantics describes how a process $P$ evolves to a process $P'$ in some context, sometimes referred to as its *observer*. An environmental semantics explicitly records the set $\Gamma$ of channels that are known to the observer. More precisely, the operational semantics takes the form of a relation

$$\langle \Gamma\,;\, P\rangle \xrightarrow{\alpha} \langle \Gamma'\,;\, Q\rangle$$

which states that $P$ can evolve to $Q$ by doing an action $\alpha$ in an context that knows $\Gamma$. The relation also describes the set $\Gamma'$ of names known by the observer after the action has taken place. The full environmental semantics are shown in Fig. 1. Actions take one of the following forms:

- $c?n$ the process inputs $n$ from its context on channel $c$
- $c!n$ the process outputs $n$ on channel $c$
- $(n)c!n$ the process outputs a *fresh* channel $n$ on channel $c$
- $\tau$ the process takes an internal transition (not involving the observer)

Process `nil` can take no transitions; $c!n.P$ is a process that can output $n$ on channel $c$ and become $P$ (Rule OUT); after this transition the observer certainly knows $n$. Process $c?x.P$ can input $n$ from the observer on channel $c$ and become $P$ with $x$ replaced by $n$ (Rule INP). The observer can pick $n$ from its knowledge environment $\Gamma$ or send a fresh channel, but in both cases $n$ will be in the knowledge environment after the action. A choice over countable number of processes is denoted by $\sum_{k\in\mathbb{N}} P_k$ and is resolved when one of the processes in the choice takes a transition (Rule SUM).

One of the important features of the $\pi$-calculus is the ability to *restrict* the scope of a channel $n$ in process $P$ with the syntax $\nu n.P$, and dynamically extend this scope via communication (*extrusion*). Process $\nu n.P$ can do all $\alpha$ actions that $P$ can do, as long as $n$ does not appear in $\alpha$; since $n$ is locally bound it must be different from the channels in $\Gamma$ and $\Gamma'$. If process $P$ outputs a locally bound channel $n$ on another channel $c$ then $\nu n.P$ takes a transition labelled $(n)c!n$ (Rule OPEN). With this transition the binder disappears around the process and $n$ is added in the knowledge of the observer; rule PAR makes sure this channel is different from the channels in the surrounding parallel processes.

Communication over a channel $c$ between two processes running in parallel $(P \mid Q)$ is achieved by combining complementary input and output transitions into a $\tau$-transition (Rule COMM-L). In this case $P$ and $Q$ are each other's observers which is why we must extend the knowledge environments in the premises of the rule. The observer of the entire process, however, sees only a $\tau$-transition and does not "learn" $c$. In the case where the output transition is an extrusion, the binder $\nu n$ is reintroduced around the parallel processes (Rule CLOSE-L).

Replication $(*P)$ allows us to write processes with infinite consecutive transitions; $*P$ takes an $\alpha$ transition when a finite number of its unfoldings take that transition (Rule REP).

*Example 1.* The following is a process that repeatedly inputs the name of a channel $x$ and outputs a fresh name on that channel:

$$\left(P \stackrel{\mathsf{def}}{=} *(\nu n.c?x.x!n.\mathtt{nil})\right) \xrightarrow{c?y_1} P \mid \nu n.y_1!n.\mathtt{nil} \xrightarrow{(n_1)y_1!n_1} P \xrightarrow{c?y_2}\xrightarrow{(n_2)y!n_2} P \dots$$

The infinite traces in this example extrude an infinite number of fresh names; such traces are important to behavioural theories such as must-testing [6] (see [12, Sec. 5] for examples).

## 3 Permutation Types

A permutation type is a type equipped with an operation to apply a permutation. The main definitions and results are summarized in Figure 2, which follows the presentation in [11]. Proofs for all the claims in the figure can also be found in the Coq sources accompanying this paper.

The definition of *support* of a term is powerful if a bit mystifying. When the support of a term is a finite set, it is justified to think of the support as the set of free names in the term. However, when the support is infinite this is no longer

A **permutation** $p$ is a bijection $\texttt{var} \to \texttt{var}$ s.t. the set $\{a \mid pa \neq a\}$ is finite.

- The identity function $\texttt{id}$ is a permutation.
- If $p_1, p_2$ are permutations then so are $p_1 \circ p_2$ and $p_1^{-1}$
- Swapping two variables, denoted $(a_1 a_2)$, is a permutation.

$$(a_1 a_2)\ a = (\text{if } a = a_1 \text{ then } a_2 \text{ else if } a = a_2 \text{ then } a_1 \text{ else } a)$$

A **permutation type** is a set $T$ equipped with an operation $p \bullet T$ (read "$p$ applied to $T$") s.t. for all $t : T$ we have $\texttt{id} \bullet t = t$ and $(p_1 \circ p_2) \bullet t = p_1 \bullet (p_2 \bullet t)$.

- Variables and permutations form permutation types.
- If $T_1, T_2$ are permutation types then so are $T_1 * T_2$, $T_1 + T_2$, $\texttt{list}\ T_1$, $\texttt{set}\ T_1$ and $T_1 \to T_2$ (where $p \bullet f : T_1 \to T_2 = p \circ f \circ p^{-1}$)
- $\emptyset$, $\texttt{unit}$, $\texttt{bool}$ and $\mathbb{N}$ are (degenerate) permutation types.

A function is **equivariant** iff $\forall p \cdot p \bullet f = f$.

- An $n$-ary function $f$ is equivariant if $p \bullet (f\ x_1\ \cdots\ x_n) = f\ (p \bullet x_1)\ \cdots\ (p \bullet x_n)$.
- An $n$-ary predicate $P$ is equivariant if $(P\ x_1\ \cdots\ x_n)$ iff $(P\ (p \bullet x_1)\ \cdots\ (p \bullet x_n))$

The **support** of a term $t : T$ is given by $\texttt{supp}(t) = \{a \mid \texttt{infinite}\ \{b \mid (ab)t \neq t\}\}$.
We write $a \# t$ for $a \notin \texttt{supp}(t)$.

**Fig. 2.** Permutation Types

the case and it is instructive to consider counter-examples. In the remainder of this section we therefore give examples of and lemmas characterizing the support for specific kinds of terms. Type variables mentioned in this section are assumed to refer to permutation types. We give proofs only where we feel they serve an educational purpose; proofs of all lemmas can be found in the Coq sources.

### 3.1 Distributing supp over constructors

**Lemma 2.** *Let $f : T_1 \to T_n \to T'$ be injective and equivariant. Then* $\texttt{supp}(f\ x_1\ \cdots\ x_n) = \texttt{supp}(x_1) \cup \cdots \cup \texttt{supp}(x_n)$.

*Proof.* After unfolding definitions, the proof is given by

$$\begin{array}{lr} \text{inf. } \{b \mid (ab) \bullet (f\ x_1\ \cdots\ x_n) \neq f\ x1\ \cdots\ x_n\} & \\ \text{iff inf. } \{b \mid (f\ ((ab) \bullet x_1)\ \cdots\ ((ab) \bullet x_n)) \neq f\ x1\ \cdots\ x_n\} & \text{(equivariance)} \\ \text{iff inf. } \{b \mid (ab) \bullet x_1 \neq x_1 \vee \cdots \vee (ab) \bullet x_n \neq x_n\} & \text{(injectivity)} \end{array}$$ □

This lemma is tremendously useful when proving properties of the support for an inductively defined datatype. For instance, $\texttt{supp}(P \mid Q) = \texttt{supp}(P) \cup \texttt{supp}(Q)$ follows immediately: all constructors of an inductive datatype are injective by definition, and equivariance follows from the definition of applying a permutation to a term. In Coq, we provide a number of lemmas corresponding to Lemma 2, one for each arity of the function, and a tactic that picks the right one depending on the goal. Proofs such as the one about the characterization of the support of parallel processes are solved fully automatically by the tactic.

### 3.2 Support of finite and infinite sets

For non-injective functions the situation is not so straightforward. For example, consider set union. We certainly have that if $S_1 = S'_1$ and $S_2 = S'_2$ then $S_1 \cup S'_1 = S_2 \cup S'_2$, which is why we have the following lemma.

**Lemma 3.** *Let $S, S'$ be (potentially infinite) sets of terms of type $T$. Then* $\mathtt{supp}(S \cup S') \subseteq \mathtt{supp}(S) \cup \mathtt{supp}(S')$.

The converse is not true. Let us consider a counter-example. Since the set of variables is countable, there must be an enumeration $x_0, x_1, \ldots$. Let us refer to the set $\mathit{Even} = \{x_0, x_2, \ldots\}$ as the "even variables" and to the set $\mathit{Odd} = \{x_1, x_3, \ldots\}$ as the "odd variables". Then $x_0 \in \mathtt{supp}(\mathit{Odd})$, *not* because $x_0$ is a free variable in $\mathit{Odd}$ but because there are an infinite number of variables $x'$ (all the odd variables) such that $(x_0 x')\mathit{Odd} \neq \mathit{Odd}$. However, $\mathtt{supp}(\mathit{Odd} \cup \mathit{Even}) = \emptyset$ because $\mathit{Odd} \cup \mathit{Even}$ is the set of *all* names, and since sets are unordered swapping two elements of a set is an identity operation. However, for finite sets containing elements with finite support the converse *is* true:

**Lemma 4.** *Let $S, S'$ be finite sets of terms with finite support. Then* $\mathtt{supp}(S \cup S') = \mathtt{supp}(S) \cup \mathtt{supp}(S')$.

### 3.3 Support of indexed sets

For indexed sets too one direction is easy, but the other direction is not in general true.

**Lemma 5.** *Let $S : \mathbb{N} \to T$ be an indexed set. Then* $\bigcup_{n \in \mathbb{N}} \mathtt{supp}(S_n) \subseteq \mathtt{supp}(S)$.

The converse is not true. Consider again the enumeration of the variables from the previous section, and the indexed set $P = [x_1, x_2, \ldots]$. Then $x_0 \in \mathtt{supp}(P)$, *not* because $x_0$ is a free variable in $P$, but because there are an infinite number of variables $x'$ such that $(x_0 x')P \neq P$ (all of $x_1, x_2, \ldots$). However, if the support of $S$ is finite, then the converse *is* true:

**Lemma 6.** *Let $S : \mathbb{N} \to T$ be an indexed set such that $\mathtt{supp}(S)$ is finite. Then* $\bigcup_{n \in \mathbb{N}} \mathtt{supp}(S_n) = \mathtt{supp}(S)$.

Unfortunately, finiteness of the support of the indexed set does not follow from finiteness of support of its elements, as the example above showed.

### 3.4 Infinite support

In the previous two sections we have explained why $x_1 \in \mathtt{supp}(\mathit{Even})$ and $x_0 \in \mathtt{supp}(P)$. But in fact we could have made stronger statements, as illustrated by the following startling lemma:

**Lemma 7.** *Let $t : T$ be some term of a permutation type. If* $\mathtt{supp}(t)$ *is infinite then* $\mathtt{supp}(t)$ *is the set of* all *names.*

The lemma follows as a corollary from Theorem 1, but before we prove that theorem we need an intermediate result about support:

**Lemma 8.** *If $a \# t$ and $b \in \mathtt{supp}(t)$ then $(ab)t \neq t$.*

We will also use the following useful characterization of infinity:

**Lemma 9 (Infinity Principle).** *A set $S$ is infinite iff for all finite sets $S'$ there exists an $x$ in $S$ but not in $S'$.*

We can now prove the following theorem:

**Theorem 1.** *If $a \# t$ then* $\mathtt{supp}(t)$ *is finite.*

*Proof (by contradiction).* Assume $a \# t$ and $\mathtt{supp}(t)$ infinite. We will show $a \in \mathtt{supp}(t)$. By unfolding definitions it suffices to show that $\{b \mid (ab) \bullet t \neq t\}$ is infinite. By Lem. 9, picking a finite set $S'$, we need to show $\exists b {\cdot} b \notin S' \wedge (ab) \bullet t \neq t$. By the same lemma, pick $b \notin S'$ and $b \in \mathtt{supp}(t)$. We now need to show $(ab) \bullet t \neq t$ which follows from Lem. 8. □

Theorem 1 can often be useful to establish that the support of a term is finite, so that results such as Lemmas 4 and 6 from the previous sections apply. The following lemma is useful for the same reason:

**Lemma 10.** *Let $f : T_1 \to \cdots \to T_n \to T'$ be an equivariant function. If* $\mathtt{supp}(t_1), \ldots, \mathtt{supp}(t_n)$ *are finite then* $\mathtt{supp}(f\ t_1\ \cdots\ t_n)$ *is finite.*

*Proof (by contradiction).* Assume $\mathtt{supp}(t_1), \ldots, \mathtt{supp}(t_n)$ finite and $\mathtt{supp}(f\ t_1\ \cdots\ t_n)$ infinite. Use Lemma 9 to pick an $a, b \# t_1, \ldots, t_n$ such that $(ab) \bullet (f\ t_1\ \cdots\ t_n) \neq (f\ t_1\ \cdots\ t_n)$. Since $f$ is equivariant and $a, b \# t_1, \ldots, t_n$ we have
$(ab) \bullet (f\ t_1\ \cdots\ t_n) = f\ ((ab)t_1)\ \cdots\ ((ab)t_n) = f\ t_1\ \cdots\ t_n$. □

## 4 Locally Nameless Permutation Types

In the locally nameless representation of terms [1] bound variables are represented by de Bruijn indices and free variables are represented by constants. The $\pi$-calculus term $\nu c.n!c$ would be represented as $\nu.n!0$. In this representation alpha-equivalent terms are identified but we avoid reasoning about indices.

We will only work with terms that are *locally closed*: all de Bruijn indices must point to a valid binder, ruling out terms such as $\nu.n!1$. When we have a term such as $\nu P$ and want to say something about $P$ (like in rule Res) we need to *open* $P$: replace bound variable 0 by a free name $x$. In general, we need an operation $\{i \triangleright x\}t$ that replaces bound index $i$ by variable $x$ in $t$, and write $t^x$ for the special case $\{0 \triangleright x\}t$. It is sometimes also useful to have the opposite operation $\{i \triangleleft x\}t$ which replaces variable $x$ by binder $i$.

For technical reasons it is useful to have two equivalent definitions of local closure. They differ principally in how they deal with binders. The main definition is simply called *local closure*, denoted `lc`. A term such as $\nu P$ is locally closed if $P^x$ is locally closed for all constants $x$ not in some finite set $L$. This definition is extremely useful when proving lemmas of the form $(\forall t, \mathtt{lc}(t) \to P(t))$ for some property $P$ of locally closed terms. If we attempted to prove this by induction on $t$ then in the case for $\nu P$ we would get an induction hypothesis about $P$, which is not locally closed. If instead we do induction on $\mathtt{lc}(t)$ we get an induction hypothesis for $P^x$, which is usually what we need. We come back to the use of cofinite quantification (for all $x$ not in a finite set $L$) in Section 5.

The alternative definition of local closure is called *local closure at level $i$*, denoted $\mathtt{lc}_i$. A term $\nu P$ is locally closed at level $i$ if $P$ is locally closed at level $i+1$. This definition is primarily useful when proving properties about the locally nameless operations (opening and closing) themselves.

Figure 3 gives our definition of a *locally nameless permutation type* (LNPT): a permutation type equipped with variable opening and closing and the two definitions of local closure, satisfying a number of axioms describing each of the operations. Figure 4 gives the locally nameless representation for the $\pi$-calculus, the implementation of variable opening and closing, and both definitions of local closure. We omit the proofs that these definitions satisfy the necessary axioms, as they are straightforward (see the Coq sources for details). The only slight complication is that the proof that $\mathtt{lc}_0$ implies `lc` requires ordinal induction due to the countable choice in the language. In the remainder of this section we discuss where in these axioms or their proofs we need the assumption of finite support.

We explicitly added the assumption of finite support only to axiom `close_supp`. Recall from Section 3.3 that $x_0$ is in the (infinite) support of the indexed set $Pos = [x_1, x_2, \ldots]$. Closing $Pos$ with respect to $c_0$ clearly is an identity operation because $c_0$ does not occur in $Pos$. Hence, it is not true that $x_0 \# \{i \triangleleft c_0\}P$. This is important because we are interested in languages with countable choice.

For the remaining lemmas we did not add an explicit assumption about finite support because it can be derived where required. For instance, consider proving axiom `open_supp` for a term $S : \mathbb{N} \to T$ (an indexed set) for some locally nameless permutation type $T$. Since $T$ is an LNPT we know that

$$\text{If } x \# S_n \text{ and } x \neq y \text{ then } x \# \{i \triangleright y\}S_n$$

for all elements $S_n$ of $S$. We are given $x \# S$ and hence $x \# S_n$ for all $n$, but unfortunately it does *not* follow from $x \# \{i \triangleright y\}S_n$ for all $n$ that $x \# \{i \triangleright y\}S$, *unless* the support of $\{i \triangleright y\}S$ is finite (Lemma 6). However, since $x \# S$ we know by Theorem 1 that the support of $S$ must be finite, and hence by Lemma 10 the support of $\{i \triangleright y\}S$ must be finite since opening is an equivariant operation.

Finally, it is worth pointing out that the original papers about the LNCQ approach mention the possibility of introducing two alternative definitions of local closure (`lc` and $\mathtt{lc}_i$) but advocate using only `lc`. In the absence of $\mathtt{lc}_i$ lemmas `open_id` and `open_close` take a premise $\mathtt{lc}(t)$ instead. The proofs of A **locally nameless permutation type** ("LN type") is a permutation type equipped with two equivariant operations $\{i \triangleright x\}t$ and $\{i \triangleleft x\}t$ (variable opening and closing) and two equivariant predicates $\texttt{lc}$ and $\texttt{lc}_i$ such that

**(Relation between $\texttt{lc}$ and $\texttt{lc}_i$)**

$\texttt{lc}_i(\{i \triangleright x\}t)$ iff $\texttt{lc}_{i+1}(t)$ (lcat_open)

$\texttt{lc}(t)$ iff $\texttt{lc}_0(t)$ (lc_iff_lc_at)

**(Properties of opening)**

If $\texttt{lc}_i(t)$ and $i \leq i'$ then $\{i' \triangleright x\}t = t$ (open_id)

If $x \# t$ and $x \neq y$ then $x \# \{i \triangleright y\}t$ (open_supp)

**(Properties of closing)**

If $x \# t$ then $\{i \triangleleft x\}t = t$ (close_id)

If $\texttt{supp}(t)$ is finite then $x \# \{i \triangleleft x\}t$ (close_supp)

**(Relation between opening and closing)**

If $x \# t$ then $\{i \triangleleft x\}\{i \triangleright x\}t = t$ (close_open)

If $\texttt{lc}_i(t)$ and $i \leq i'$ then $\{i' \triangleright x\}\{i' \triangleleft x\}t = t$ (open_close)

If $T_1, T_2$ are LN types then so are $T_1 * T_2$, $T_1 + T_2$, $\texttt{list}\ T_1$ and $\mathbb{N} \rightarrow T_1$.

**Fig. 3.** Locally Nameless Permutation Types

these lemmas then rely on two technical lemmas which must be added to the set of axioms defining a locally nameless type:

$$\text{If } i \neq j \text{ and } \{i \triangleright x\}\{j \triangleright y\}t = \{j \triangleright y\}t \text{ then } \{i \triangleright x\}t = t$$
$$\text{If } i \neq j, x \neq z \text{ and } x \# t \text{ then } \{i \triangleright x\}\{j \triangleright y\}\{j \triangleleft z\}t = \{j \triangleright y\}\{j \triangleleft z\}\{i \triangleright x\}t$$

The authors of the LNCQ papers prefer introducing these technical lemmas at the benefit of not needing an additional local closure property, but we feel that the presentation using $\texttt{lc}_i$ leads to a more natural axiomatization. Moreover, this presentation has the additional benefit that in the proof of axiom open_close we do not need to reason about finiteness of support.

## 5 Cofinite Quantification

The naive translation of the rule for restriction (Fig. 1) to locally nameless style is

$$\frac{\langle \Gamma\,;\, P^c \rangle \xrightarrow{\alpha} \langle \Gamma'\,;\, Q^c \rangle \qquad c \# \alpha, \Gamma, \Gamma', P, Q}{\langle \Gamma\,;\, \nu P \rangle \xrightarrow{\alpha} \langle \Gamma'\,;\, \nu Q \rangle}\ \textsc{Res-Ex}$$

The authors of [1] call this style of rule "exists-fresh", and explain that although this rule provides a convenient introduction form (the premises of the rule only have to be shown for a single fresh variable), it only provides a weak induction

**Syntax**

$$n \in \texttt{Name} ::= x \mid i$$

$$P, Q \in \texttt{Term} ::= \texttt{nil} \mid \sum [P]^{\infty} \mid n?P \mid n!m.P \mid (P \mid Q) \mid \nu P \mid *P$$

**Opening and Closing**

$$\begin{aligned}
\{i \triangleright x\}y &= y \\
\{i \triangleright x\}j &= \begin{cases} x & \text{if } i = j \\ j & \text{otherwise} \end{cases}
\end{aligned}
\qquad
\begin{aligned}
\{i \triangleleft x\}y &= \begin{cases} i & \text{if } x = y \\ y & \text{otherwise} \end{cases} \\
\{i \triangleleft x\}j &= j
\end{aligned}$$

$$\begin{aligned}
\{i \triangleright x\}(\texttt{nil}) &= \texttt{nil} \\
\{i \triangleright x\}(\textstyle\sum S) &= \textstyle\sum \{i \triangleright x\}S \\
\{i \triangleright x\}(n?P) &= \{i \triangleright x\}n?\{i+1 \triangleright x\}P \\
\{i \triangleright x\}(n!m.P) &= \{i \triangleright x\}n!\{i \triangleright x\}m.\{i \triangleright x\}P \\
\{i \triangleright x\}(P \mid Q) &= \{i \triangleright x\}P \mid \{i \triangleright x\}Q \\
\{i \triangleright x\}(\nu P) &= \nu\{i+1 \triangleright x\}P \\
\{i \triangleright x\}(*P) &= *\{i \triangleright x\}P
\end{aligned}
\qquad
\begin{aligned}
\{i \triangleleft x\}(\texttt{nil}) &= \texttt{nil} \\
\{i \triangleleft x\}(\textstyle\sum S) &= \textstyle\sum \{i \triangleleft x\}S \\
\{i \triangleleft x\}(n?P) &= \{i \triangleleft x\}n?\{i+1 \triangleleft x\}P \\
\{i \triangleleft x\}(n!m.P) &= \{i \triangleleft x\}n!\{i \triangleleft x\}m.\{i \triangleleft x\}P \\
\{i \triangleleft x\}(P \mid Q) &= \{i \triangleleft x\}P \mid \{i \triangleleft x\}Q \\
\{i \triangleleft x\}(\nu P) &= \nu\{i+1 \triangleleft x\}P \\
\{i \triangleleft x\}(*P) &= *\{i \triangleleft x\}P
\end{aligned}$$

**Local Closure**

*Names*

$$\frac{}{\texttt{lc}(x)}\ \text{LC-Free}$$

*Processes*

$$\frac{\texttt{lc}(Ps)}{\texttt{lc}(\sum Ps)}\ \text{LC-Sum} \qquad \frac{\texttt{lc}(P) \quad \texttt{lc}(Q)}{\texttt{lc}(P \mid Q)}\ \text{LC-Par}$$

$$\frac{\texttt{lc}(n) \quad (\forall x \notin L \cdot \texttt{lc}(P^x))}{\texttt{lc}(n?P)}\ \text{LC-Inp} \qquad \frac{\texttt{lc}(n) \quad \texttt{lc}(m) \quad \texttt{lc}(P)}{\texttt{lc}(n!m.P)}\ \text{LC-Out}$$

$$\frac{(\forall x \notin L \cdot \texttt{lc}(P^x))}{\texttt{lc}(\nu P)}\ \text{LC-Res} \qquad \frac{\texttt{lc}(P)}{\texttt{lc}(*P)}\ \text{LC-Rep} \qquad \frac{}{\texttt{lc}(\texttt{nil})}\ \text{LC-Nil}$$

**Local Closure at Level $i$**

*Names*

$$\frac{}{\texttt{lc}_i(x)}\ \text{LC}_i\text{-Free} \qquad \frac{i' < i}{\texttt{lc}_i(i')}\ \text{LC}_i\text{-Bound}$$

*Processes*

$$\frac{\texttt{lc}_i(Ps)}{\texttt{lc}_i(\sum Ps)}\ \text{LC}_i\text{-Sum} \qquad \frac{\texttt{lc}_i(P) \quad \texttt{lc}_i(Q)}{\texttt{lc}_i(P \mid Q)}\ \text{LC}_i\text{-Par}$$

$$\frac{\texttt{lc}_i(n) \quad \texttt{lc}_{i+1}(P)}{\texttt{lc}_i(n?P)}\ \text{LC}_i\text{-Inp} \qquad \frac{\texttt{lc}_i(n) \quad \texttt{lc}_i(m) \quad \texttt{lc}_i(P)}{\texttt{lc}_i(n!m.P)}\ \text{LC}_i\text{-Out}$$

$$\frac{\texttt{lc}_{i+1}(P)}{\texttt{lc}_i(\nu P)}\ \text{LC}_i\text{-Res} \qquad \frac{\texttt{lc}_i(P)}{\texttt{lc}_i(*P)}\ \text{LC}_i\text{-Rep} \qquad \frac{}{\texttt{lc}_i(\texttt{nil})}\ \text{LC}_i\text{-Nil}$$

**Fig. 4.** The $\pi$-calculus as a LNPT

principle. They give an example of a failed proof of *weakening* of a typing relation. The same issue arises here; indeed, we chose an environmental semantics for our study precisely to be able to give a similar example (but similar issues arise when using other styles of transition semantics). Consider:

**Lemma 11 (Weakening).** *If* $\langle \Gamma\,;\,P\rangle \xrightarrow{\alpha} \langle \Gamma'\,;\,Q\rangle$ *then for all* $\Gamma''$ *such that* $\mathsf{extr}(\alpha) \cap \Gamma'' = \emptyset$ *we have* $\langle \Gamma, \Gamma''\,;\,P\rangle \xrightarrow{\alpha} \langle \Gamma', \Gamma''\,;\,Q\rangle$*.*

The side condition $\mathsf{extr}(\alpha) \cap \Gamma'' = \emptyset$ makes sure that the "fresh" names extruded by the process are different from the additional environment. Unfortunately, we cannot prove this lemma using rule Res-Ex.

*Proof (First failed attempt).* By induction on the transition relation. In the case for restriction, we get an induction hypothesis at $P^c$ for a single $c$. Unfortunately, we do not know if $c \notin \Gamma''$ so the proof fails (note that equivariance of the transition relation here does not help). $\times$

The style of presentation promoted in the LNCQ approach is *cofinite quantification.* For restriction this translates to

$$\frac{\forall n \notin L \cdot \left(\langle \Gamma\,;\,P^n\rangle \xrightarrow{\alpha} \langle \Gamma'\,;\,Q^n\rangle\right)}{\langle \Gamma\,;\,\nu P\rangle \xrightarrow{\alpha} \langle \Gamma'\,;\,\nu Q\rangle}\,\textsc{Res}$$

Instead of requiring the premise for a single $n$, we require it for *all* $n$ not in a finite set $L$. This rules gives a much stronger induction hypothesis but is almost as convenient as an introduction form because we can exclude any set of names that might cause a clash. As a bonus, we do not have to be so precise anymore about what $c$ needs to be fresh from. Cofinite quantification gives us a mathematically precise version of the ever popular side-condition "$c$ sufficiently fresh". With the new rule the case for restriction in the weakening proof becomes immediate.

So far we have simply followed the recommendations given by the LNCQ approach, but the situation is less clear-cut for some other rules. For instance, do we need cofinite quantification in the rule for extrusion?

$$\frac{\forall n \notin L \cdot \left(\langle \Gamma\,;\,P^n\rangle \xrightarrow{c!n} \langle \Gamma, n\,;\,Q^n\rangle\right) \quad m \# c, Q, \Gamma \quad m \notin L}{\langle \Gamma\,;\,\nu P\rangle \xrightarrow{(m)c!m} \langle \Gamma, m\,;\,Q^m\rangle}\,\textsc{Open-CQ}$$

As it turns out, we do not. Cofinite quantification does not help when using this rule as an introduction form because we cannot use $L$ to exclude $m$. Moreover, we do not need the stronger induction principle because freshness conditions about $m$ must be stated in the proof context anyway. The side condition in the weakening lemma is a prime example: the rule for restriction without cofinite quantification poses no difficulty in the weakening proof because the side condition ensures that the name we picked is not in $\Gamma''$. The question now is whether we need cofinite quantification in any other rules. Surprisingly, we do. As it stands, the proof of weakening gets stuck in the rule for closing.

*Proof (Lemma 11, Second failed attempt).* By induction on the transition relation. Consider the case for closing.

$$\frac{\langle \Gamma, c\,;\, P\rangle \xrightarrow{(n)c!n} \langle \Gamma, c, n\,;\, P'^{n}\rangle \qquad \langle \Gamma, c, n\,;\, Q\rangle \xrightarrow{c?n} \langle \Gamma, c, n\,;\, Q'^{n}\rangle}{\langle \Gamma\,;\, P \mid Q\rangle \xrightarrow{\tau} \langle \Gamma\,;\, \nu(P' \mid Q')\rangle}\ \textsc{Close-L-Ex}$$

The side condition to the lemma tells us that the extruded names in the action ($\tau$) are disjoint from $\Gamma''$. Since a $\tau$ action does not extrude any names, this is vacuously true. To be able to apply the induction hypothesis at the left (extrusion) premise we need to know that the extruded name $n$, sent from the left process to the right, is disjoint from $\Gamma''$. We do not know this, nor can we strengthen the side condition to the weakening lemma to exclude this case, as channel $n$ is an internal communication. We are once again stuck. ×

In an informal proof we would gloss over this issue; the argument would be along the lines that "clearly, $n$ is a fresh name, so it must be different from any free name in the environment". However, since there are no binders in this rule, this is difficult to make more precise. The Barendregt convention does not help, since the name is no longer bound, and moreover we have to be careful with the Barendregt convention in the presence of extrusion (Section 1). Fortunately, cofinite quantification comes to the rescue. The rule we adopt for closing is

$$\frac{\forall n \notin L \cdot \left(\langle \Gamma, c\,;\, P\rangle \xrightarrow{(n)c!n} \langle \Gamma, c, n\,;\, P'^{n}\rangle \qquad \langle \Gamma, c, n\,;\, Q\rangle \xrightarrow{c?n} \langle \Gamma, c, n\,;\, Q'^{n}\rangle\right)}{\langle \Gamma\,;\, P \mid Q\rangle \xrightarrow{\tau} \langle \Gamma\,;\, \nu(P' \mid Q')\rangle}\ \textsc{Close-L}$$

As before, the intuitive reading of the cofinite quantification in the rule is that the left process must extrude a fresh name, and the right process must input a fresh name. This is of course exactly the intended meaning of this rule. With the modification, the proof of weakening is straightforward (the full operational semantics in locally nameless style is shown in Figure 5).

*Proof (Lemma 11).* By induction on the transition relation. In the case for closing we apply rule Close-CQ, instantiating $L$ to $L' \cup \Gamma''$ (where $L'$ is the set we got from the rule induction). □

This proof is almost as easy as the pencil-and-paper proof (the entire weakening proof only takes a few lines in Coq), and we do not need to reason about renaming at all. In particular, in the case for the rule for closing, we can simply assume that the communicated channel is disjoint from the new environment, just like in the pencil-and-paper proof. This of course is the primary purpose of using cofinite quantification. However, it is not at all obvious *a priori* that we should use cofinite quantification in the rule for closing, which after all does not talk about any bound names.[2]

[2] Alternatively, we can prove the lemma by induction on the *height* of the derivation by taking advantage of equivariance. The resulting proof is however more cumbersome; moreover, induction on height is awkward in a theorem prover such as Coq.

$$\frac{}{\langle \Gamma, c\,;\, c?P\rangle \xrightarrow{c?n} \langle \Gamma, c, n\,;\, P^n\rangle}\ \textsc{Inp}$$

$$\frac{\forall n \notin L \cdot \left(\langle \Gamma\,;\, P^n\rangle \xrightarrow{\alpha} \langle \Gamma'\,;\, Q^n\rangle\right)}{\langle \Gamma\,;\, \nu P\rangle \xrightarrow{\alpha} \langle \Gamma'\,;\, \nu Q\rangle}\ \textsc{Res} \qquad \frac{\langle \Gamma\,;\, P^n\rangle \xrightarrow{c!n} \langle \Gamma, n\,;\, Q^n\rangle \quad n \# P, Q, \Gamma}{\langle \Gamma\,;\, \nu P\rangle \xrightarrow{(n)c!n} \langle \Gamma, n\,;\, Q^n\rangle}\ \textsc{Open}$$

$$\frac{\forall n \notin L \cdot \left(\langle \Gamma, c\,;\, P\rangle \xrightarrow{(n)c!n} \langle \Gamma, c, n\,;\, P'^n\rangle \qquad \langle \Gamma, c\,;\, Q\rangle \xrightarrow{c?n} \langle \Gamma, c, n\,;\, Q'^n\rangle\right)}{\langle \Gamma\,;\, P \mid Q\rangle \xrightarrow{\tau} \langle \Gamma\,;\, \nu(P' \mid Q')\rangle}\ \textsc{Close-L}$$

**Fig. 5.** Operational Semantics in locally nameless style (changed rules only)

## 6 The Need for Equivariance

Although cofinite quantification eliminated the need to reason about equivariance completely in the proof about weakening, the presence of an extrusion rule reduces its effectiveness in other proofs. For instance, consider

**Lemma 12.** *If* $\langle \Gamma\,;\, P\rangle \xrightarrow{\alpha} \langle \Gamma\,;\, Q\rangle$ *and* $\texttt{lc}(P)$ *then* $\texttt{lc}(Q)$.

*Proof.* By induction on the transition relation. Consider the case for extrusion. The induction hypothesis tells us that if $P^c$ is locally closed then $Q^c$ is locally closed (for one specific $c$). Since $\texttt{lc}(\nu P)$ we know that $P^x$ is locally closed for all $x$ not in some set $L$. Since we do not know if $c \notin L$ we cannot use this directly.

Instead, we need to take advantage of equivariance of `lc`. Pick a name $x \# L, P$. Local closure of $\texttt{lc}(\nu P)$ gives us that $P^x$ is locally closed. Moreover, since $c, x \# P$ and since opening is equivariant, this means that $P^c$ is locally closed and we can finish the proof by applying the induction hypothesis. □

This proof relies on the following simple but useful lemma:

**Lemma 13.** *Given equivariant property* $P$ *and* $x, y \# t$*, if* $P(t^x)$ *then* $P(t^y)$.

Note that the presence of extrusion in the *transition* relation made it necessary to take advantage of equivariance of *another* relation (in this case, local closure). If we had a typing relation on processes and tried to prove that typing was preserved by the transition relation we would need to take advantage of equivariance of the typing relation. Therefore reasoning about equivariance is more prevalent in a language with extrusion than it is in a language such as the $\lambda$-calculus. For this reason we based our locally nameless types on permutation types, which excel at reasoning about renaming and make it possible to prove numerous lemmas about equivariance and support without reference to the specific syntax of the language.

Note that we still need to rely on equivariance when using a cofinite version of the extrusion rule. In that case, the induction hypothesis would tell us that for all $x$ not in some finite set $L$ if $P^x$ is locally closed then $Q^x$ is locally closed.

Inversion on local closure of $\nu P$ tells us that $P^x$ is locally closed for all $x$ in some *other* set $L'$. We need to show that $Q^c$ is locally closed, where we know that $c \notin L$. Hence, we can apply the induction hypothesis, leaving us to to show that $P^c$ is locally closed. Unfortunately, we do not know if $c \notin L'$ so again we need to rely on equivariance of local closure to finish the proof. The only difference between this proof and the proof using the version of the rule without cofinite quantification is *when* we need to reason about equivariance: before or after applying the induction hypothesis.

Finally, equivariance of the transition relation itself is an important property in its own right. In particular, it has the following immediate corollary:

**Corollary 1.** *If $t$ is a trace of $P$ then for all $n \# P$ and $m \# P$ the trace $(n m) t$ is a trace of $P$.*

This ability to rename the "new" (input or extruded) names in a—potentially infinite—trace is often important in proofs about process calculi [6].

## 7 Related Work and Conclusions

The treatment of names is one of the first stumbling blocks towards formalizing programming language theory. Process calculi often contain rules in which bound names become free, which complicates the treatment of names further still. We have discussed extrusion in detail, but it also arises in other contexts. For instance, our work was motivated by the formalization of the translation of a language with binders to a language without [21].

In this paper we proposed an approach that elegantly deals with such formalizations. Our approach combines two prominent approaches to dealing with names: the locally nameless approach, first used by Gordon [7] and popularized and extended with the use of cofinite quantification by Aydemir et al. [1, 5], and the nominal approach [16, 19]. We have already discussed both approaches extensively. A detailed survey of other approaches to dealing with name binding is beyond the scope of this paper; we focus on formalizations of the $\pi$-calculus.

Early work on the mechanical formalization of the $\pi$-calculus used the naive approach for names [13, 15] with little success, or the de Bruijn approach [9] with the considerable burden of proving lemmas about indices.

Some more recent formalizations used higher-order abstract syntax (HOAS) in Coq and Isabelle [10, 17]. This approach involves proving technical results that have little to do with the pencil-and-paper proofs (e.g. about "exotic" terms). Moreover certain properties of substitution and freshness of names are not directly provable and need to be postulated.

Perhaps the approach in formalizing $\pi$-calculus that was most faithful to handwritten proofs was that of Bengtson and Parrow [3, 4] in Nominal Isabelle. As we discussed in the introduction, however, the nominal approach can only be used with the semantics that use concretions and not the more common Ext rule. Their work in formalizing concurrency theory is quite extensive, including a theory of bisimulation for the $\pi$-calculus. A more detailed comparison between their approach and locally closed permutation types will be possible after we use the latter for a formalization of the behavioural theory of the $\pi$-calculus, which we leave for future work.